\documentclass[conference,lettersize]{IEEEtran}

\IEEEoverridecommandlockouts
\usepackage{amsmath,amsfonts,amssymb}
\usepackage{graphicx}
\usepackage{booktabs}
\usepackage{algorithm}
\usepackage{algorithmic}
\usepackage{balance}
\usepackage{subcaption}

\title{Degeneracy-Aware Functional and Algorithmic Resilience in Virtualized 6G Networks Under Correlated Failures}

\author{
\IEEEauthorblockN{Mohamed Khalafalla Hassan and Indrakshi Dey}
\IEEEauthorblockA{Walton Institute, South East Technological University, Waterford, Ireland\\
Email: mohamed.khalafala.hassan@gmail.com, indrakshi.dey@waltoninstitute.ie}
}

\begin{document}
\maketitle

\begin{abstract}
Redundancy is widely used to sustain service continuity in programmable and virtualized networks; however, replicated functions often share platforms, software stacks, and control dependencies, making them vulnerable to correlated failures. Consequently, replica counts alone may overestimate true resilience. This paper adopts a degeneracy-aware perspective, where robustness depends on the availability of structurally diverse yet functionally equivalent alternatives. We formalize this perspective through three complementary metrics: the Functional Substitution Score (FSS), which quantifies structurally distinct substitutes for a function; the Algorithmic Resilience Quotient (ARQ), which measures diversity among algorithms that remain comparable in delivered performance; and the Multi-Layer Degeneracy Index (MLDI), which captures how functional diversity is distributed across architectural layers. Using targeted disruption protocols on a synthesized data, we show that redundancy and robustness can diverge substantially. The results show that FSS separates structural diversity from replica count, ARQ distinguishes genuine algorithmic alternatives from near-duplicate implementations, and MLDI captures cross-layer buffering that remains hidden under redundancy-only analysis. These findings establish degeneracy as a practical resilience primitive for open, disaggregated, and virtualized 6G systems.
\end{abstract}

%\vspace{-3mm}
\section{Introduction}

Resilience is becoming a first-order design requirement in 6G systems, where communication, computation, and control are increasingly embedded into critical services [1], [2]. The shift toward virtualization, disaggregation, and cloud-native orchestration expands deployment flexibility, but it also increases dependence on shared software stacks, orchestration frameworks, and physical substrates [3]--[5]. In such settings, disruptions are often correlated rather than independent, and the effective robustness of the system can be much lower than nominal redundancy suggests [1], [2].

This exposes a central limitation of redundancy-centric thinking. Counting replicas is informative only when the replicas fail independently. In virtualized and software-defined environments, however, multiple instances may share the same implementation lineage, orchestration stack, or control-plane dependencies, so a single disruption can invalidate several apparent backups at once [6]--[8]. What matters is not only how many alternatives exist, but whether those alternatives are sufficiently different to avoid common-mode collapse.

Degeneracy provides a principled way to capture this distinction. In this context, degeneracy refers to the ability of structurally different elements to deliver equivalent functionality under shared constraints~[9]. Unlike classical redundancy, which emphasizes replication, degeneracy emphasizes diversity in realization. In communication systems, such diversity can arise from heterogeneous software stacks, deployment locations, hardware configurations, algorithmic pipelines, or layer-specific implementations~[1,10]. The resilience value of an alternative therefore depends not only on its existence, but also on how differently it is realized. Motivated by this perspective, this paper develops a compact degeneracy-aware resilience framework built around three complementary metrics: the Functional Substitution Score (FSS), which quantifies the richness of structurally distinct substitutes for a network function; the Algorithmic Resilience Quotient (ARQ), which measures whether a task can be sustained by algorithms that remain similar in delivered performance while differing in internal structure; and the Multi-Layer Degeneracy Index (MLDI), which captures how functional diversity is distributed across the system stack and how it buffers vertical failure propagation. 

The main contributions of this paper are threefold. First, we formulate a degeneracy-aware resilience framework that separates functional equivalence from structural realization and instantiate it through FSS, ARQ, and MLDI across component, algorithmic, and cross-layer levels. Second, we evaluate these metrics under targeted disruption protocols and show that redundancy and robustness can diverge sharply. Third, we provide design-oriented interpretations showing why structural diversity, rather than replication alone, is the key enabler of graceful degradation in virtualized 6G systems.

\section{Related Work and Motivation}

Recent work increasingly treats resilience in 6G as a multi-layer property that must be engineered into the architecture rather than appended through ad hoc replication [1], [2]. Existing work has clarified the interplay among reliability, availability, safety, and security in softwarized infrastructures, but much of it remains conceptual or still relies on redundancy-based indicators such as replica count, path multiplicity, and isolated threat models. This limitation is more visible in Open RAN and cloud-native deployments. While openness and disaggregation increase implementation choice and multi-vendor flexibility, they also broaden shared vulnerabilities across open interfaces, O-Cloud platforms, CI/CD pipelines, and management stacks [6]--[8], [10]. Similarly, NFV and cloud-native studies show that orchestrators, service meshes, container images, and hypervisors induce couplings that weaken the independence assumptions behind classical reliability analysis [3]--[5], [11]. The key issue is not only whether substitutes exist, but whether they are sufficiently distinct to limit common-mode failure. This is reinforced by security and dependability studies on network slicing, cloud-native mobile cores, and next-generation cellular infrastructures, which show that modern architectures increase both flexibility and dependency density, making collapse more likely when structural diversity is overlooked [12]--[18]. The gap addressed here is the lack of a compact and computable operator for substitute diversity under correlated failures. This paper addresses that gap by focusing on three metrics that expose degeneracy across components, algorithms, and architectural layers.

\section{Degeneracy-Aware Formulation}
We define the proposed degeneracy-aware metrics as,\footnote{\textit{Notation:}
$E_f=\{e_1,\ldots,e_n\}$: set of elements realizing function $f$, $n=|E_f|$;
$D_{ij}\in[0,1]$: structural dissimilarity between $e_i$ and $e_j$;
$\delta$: structural threshold;
$I_{ij}(\delta)=\mathbf{1}\{D_{ij}>\delta\}$: distinctness indicator;
$C_i,L_i$: capacity and load of $e_i$;
$W_{ij}$: pairwise admissibility weight;
$A_i$: availability;
$R_i(T)$: reliability over mission time $T$;
$A_{\min},R_{\min}$: admissibility thresholds;
$\mathrm{MTBF}_i$: mean time between failures;
$w_i$: node-level operational weight;
$\mathrm{FSS}(f),\mathrm{FSS}^{\star(w)}(f)$: baseline and weighted Functional Substitution Scores.
$\mathcal{A}=\{A_1,\ldots,A_n\}$: algorithm set;
$P(A_i)\in\mathbb{R}^d$, $S(A_i)\in\mathbb{R}^k$: performance and structural descriptors;
$K_P(i,j)$: Gaussian performance-similarity kernel;
$\sigma$: kernel width;
$D_s(i,j)$: cosine-based structural dissimilarity;
$\epsilon$: functional similarity threshold;
$\mathrm{ARQ},\mathrm{ARQ}^{\star}$: hard and soft Algorithmic Resilience Quotients.
$\{\ell_1,\ldots,\ell_k\}$: set of layers;
$E_\ell,D_\ell$: total and admissible/distinct element sets in layer $\ell$;
$\tau_\ell=|D_\ell|/|E_\ell|$: layer-wise degeneracy ratio;
$m$: number of functions;
$p^{(\ell)}(f_i)$: fraction of layer-$\ell$ elements supporting function $f_i$;
$H(\ell)$: layer entropy;
$\mathrm{MLDI},\mathrm{MLDI}^{\star}$: baseline and entropy-enhanced Multi-Layer Degeneracy Indices;
$\gamma$: cross-layer weighting factor;
$q$: removal fraction;
L1, L2, L3: physical, control/virtualization, and service/application layers.}

\subsubsection{Functional Substitution Score (FSS)}

Let $E_f=\{e_1,\ldots,e_n\}$ denote the set of elements capable of realizing function $f$. Each element is characterized by a structural signature together with operational attributes such as capacity and load. Pairwise structural dissimilarity is represented by $D_{ij}\in[0,1]$, where larger values indicate more distinct realizations. For a structural threshold $\delta$, define $I_{ij}(\delta)=\mathbf{1}\{D_{ij}>\delta\}, \quad i\neq j$. The baseline FSS is then
\begin{equation}
\mathrm{FSS}(f)=\frac{1}{n(n-1)}\sum_{i\neq j} I_{ij}(\delta),
\end{equation}
which measures the fraction of ordered realization pairs that are structurally distinct. This baseline isolates substitute diversity from operational constraints, making it useful for distinguishing specialization from sheer replication.

To account for admissibility and substitute quality, the pairwise contribution is weighted as $W_{ij}=\frac{\min(C_i,C_j)}{1+|L_i-L_j|}D_{ij}$, where $C_i$ and $L_i$ denote capacity and instantaneous load, respectively. Operational realism is further incorporated through node weights
\begin{equation}
w_i=
\begin{cases}
1, ~~\text{None},\\
A_i\mathbf{1}\{A_i\ge A_{\min}\},~~\text{Availability},\\
R_i(T)\mathbf{1}\{R_i(T)\ge R_{\min}\},~~\text{Reliability},\\
A_iR_i(T)\mathbf{1}\{A_i\ge A_{\min}\}\mathbf{1}\{R_i(T)\ge R_{\min}\},~~\text{Joint},
\end{cases}
\end{equation}
with reliability defined as $R_i(T)=\exp\left(-\frac{T}{\mathrm{MTBF}_i}\right)$. The weighted score becomes
\begin{equation}
\mathrm{FSS}^{\star(w)}(f)=\frac{1}{n(n-1)}\sum_{i\neq j} I_{ij}(\delta)W_{ij}w_iw_j.
\end{equation}

\begin{algorithm}[h]
\caption{FSS under targeted removals}
\begin{algorithmic}[1]
\STATE Compute an importance score for each element
\STATE Rank elements in descending order of importance
\FOR{each removal fraction $q$}
    \STATE Remove the top-$q$ fraction of elements
    \STATE Recompute the structural dissimilarity matrix $D$
    \STATE Evaluate $\mathrm{FSS}^{\star(w)}$
\ENDFOR
\end{algorithmic}
\end{algorithm}

\subsubsection{Algorithmic Resilience Quotient (ARQ)}

Resilience in modern wireless systems is not determined solely by hardware or deployment structure. It also depends on the diversity of algorithms used for control, inference, optimization, and signal processing. Two algorithms may be functionally similar in delivered performance while differing substantially in structure, assumptions, and failure modes.

Let $\mathcal{A}=\{A_1,\ldots,A_n\}$ denote a set of candidate algorithms for a task. Each algorithm is represented by a performance descriptor $P(A_i)\in\mathbb{R}^d$ and a structural descriptor $S(A_i)\in\mathbb{R}^k$. The enhanced ARQ is defined as
\begin{equation}
\mathrm{ARQ}^{\star}=\frac{1}{n(n-1)}\sum_{i\neq j} K_P(i,j)D_s(i,j),
\end{equation}
where $K_P(i,j)=\exp\left(-\frac{\|P(A_i)-P(A_j)\|^2}{2\sigma^2}\right)$, and $D_s(i,j)=1-\frac{S(A_i)\cdot S(A_j)}{\|S(A_i)\|\|S(A_j)\|}$. A high ARQ indicates that the task can be sustained by algorithms that remain close in performance while differing meaningfully in internal realization. This property is critical under domain shift, model mismatch, or implementation-specific vulnerabilities, where structurally similar algorithms tend to fail together. The motivation for emphasizing structural algorithmic diversity is aligned with broader work on robustness under uncertainty and adaptive control in wireless systems [1],[2]. In a 6G setting, the performance descriptor \(P(A_i)\) can include metrics such as achieved throughput, end-to-end latency, and bit error rate under a common traffic and channel condition, whereas the structural descriptor \(S(A_i)\) can encode the underlying design and implementation lineage, for example minimum mean-squared error (MMSE) versus zero-forcing receiver logic, centralized versus distributed scheduling logic, and software stack features such as C++ versus Rust implementations or different library dependencies.

\begin{algorithm}[h]
\caption{ARQ evaluation}
\begin{algorithmic}[1]
\STATE Compute performance descriptors $P(A_i)$
\STATE Compute structural descriptors $S(A_i)$
\FOR{each algorithm pair $(i,j)$}
    \STATE Compute $K_P(i,j)$ and $D_s(i,j)$
\ENDFOR
\STATE Aggregate pairwise contributions to obtain $\mathrm{ARQ}^{\star}$
\end{algorithmic}
\end{algorithm}

\subsubsection{Multi-Layer Degeneracy Index (MLDI)}

Resilience in 6G systems is inherently cross-layer. In the proposed model, the architecture is represented by three vertically coupled layers: the physical layer $L_{1}$, the control or virtualization layer $L_{2}$, and the service or application layer $L_{3}$. Let $E_{\ell}$ denote the set of elements in layer $\ell$, and let $B^{(1,2)} \in \{0,1\}^{|E_{2}| \times |E_{1}|}$ and $B^{(2,3)} \in \{0,1\}^{|E_{3}| \times |E_{2}|}$ denote binary dependency matrices, where $B^{(1,2)}_{ji}=1$ means that control element $e^{(2)}_{j}$ depends on physical element $e^{(1)}_{i}$, and $B^{(2,3)}_{kj}=1$ means that service element $e^{(3)}_{k}$ depends on control element $e^{(2)}_{j}$. After a disruption, each element has state $x^{(\ell)} \in \{0,1\}$, and failure propagates upward because an element in $L_{2}$ or $L_{3}$ remains active only if at least one required upstream dependency remains active. Accordingly, a failure in $L_{1}$ can deactivate multiple virtualization functions in $L_{2}$ and can then reduce service availability in $L_{3}$. Degeneracy acts as a buffering mechanism when a function in $L_{3}$ retains access to multiple admissible and structurally distinct support paths through $L_{2}$ despite losses in $L_{1}$. This explicit dependency model clarifies that MLDI does not measure diversity within isolated layers only, but also the extent to which cross-layer functional continuity is preserved under vertically propagated failures.

Consider $k$ layers $\{\ell_1,\ldots,\ell_k\}$. Let $E_\ell$ denote the set of elements in layer $\ell$, and let $D_\ell\subseteq E_\ell$ denote the subset that remains functionally admissible and structurally distinct. The baseline layer-wise degeneracy ratio is $\tau_\ell = {|D_\ell|}/{|E_\ell|}$, and the baseline MLDI is 
\begin{equation}
\mathrm{MLDI}=\frac{1}{k}\sum_{\ell=1}^{k}\tau_\ell.
\end{equation}
To capture how functionality is distributed within each layer, let $p^{(\ell)}(f_i)$ denote the fraction of layer-$\ell$ elements supporting function $f_i$. The corresponding entropy is $H(\ell)=-\sum_{i=1}^{m} p^{(\ell)}(f_i)\log p^{(\ell)}(f_i)$. The entropy-enhanced variant is
\begin{equation}
\mathrm{MLDI}^{\star}=\frac{1}{k}\sum_{\ell}\frac{H(\ell)}{\log m}.
\end{equation}

This enhanced form reveals whether diversity is broadly distributed or concentrated in a few dominant elements. It therefore captures the buffering role of upper-layer diversity in the presence of lower-layer rigidity, an issue that is especially relevant in cloud-native and NFV-based systems with strong vertical coupling [3]--[5].

\begin{algorithm}[h]
\caption{MLDI computation}
\begin{algorithmic}[1]
\FOR{each layer $\ell$}
    \STATE Identify admissible and structurally distinct elements
    \STATE Compute $\tau_\ell=|D_\ell|/|E_\ell|$
    \STATE Compute functional entropy $H(\ell)$
\ENDFOR
\STATE Aggregate layer scores to obtain $\mathrm{MLDI}$ and $\mathrm{MLDI}^{\star}$
\end{algorithmic}
\end{algorithm}

\subsubsection{Computational Complexity and Intended Use}

The proposed metrics are designed for resilience assessment and decision support rather than for per-packet control. For a function with $n_f$ candidate realizations, FSS requires pairwise distinctness evaluation and therefore has $O(n_f^2)$ time complexity, with no change in asymptotic order for the weighted form in (6) because availability, reliability, and admissibility checks are linear once the pairwise matrix is available. For a portfolio of $a$ algorithms with performance descriptor dimension $d$ and structural descriptor dimension $k$, ARQ requires all pairwise kernel and dissimilarity evaluations, which gives $O(a^2(d+k))$. For MLDI, the entropy terms are linear in the number of supported functions per layer, but identifying admissible and structurally distinct substitutes across the three layers requires pairwise checks inside each layer, which yields $O\!\left(\sum_{\ell=1}^{3}|E_{\ell}|^2\right)$ in the direct implementation. Under targeted-removal experiments, these costs scale linearly with the number of removal fractions and repeated trials. Accordingly, FSS, ARQ, and MLDI are intended primarily for offline network planning, resilience benchmarking, what-if analysis, and slower supervisory orchestration loops, where structural diversity must be evaluated before or during reconfiguration. They are not intended as ultra-low-latency healing primitives inside the fastest 6G control path. In practice, however, pairwise dissimilarity matrices can be precomputed and updated incrementally, which reduces recomputation overhead and makes the metrics suitable for periodic orchestration decisions on moderate candidate sets.

\section{Results and Discussion}

The results in this section are obtained from controlled numerical experiments on synthesized deployment instances modeled on typical O-RAN and NFV deployment parameters reported in~\cite{Gao2024}. The synthetic instances are designed to reproduce the heterogeneity expected in disaggregated virtualized deployments while preserving full control over targeted-failure experiments and repeatability. In particular, node capacities are sampled from a truncated log-normal distribution to capture the right-skewed mix of edge, regional, and pooled compute resources, while node loads are sampled from a beta distribution and scaled by the assigned capacity so that the resulting utilization spans lightly loaded and moderately stressed operating points. Availability values are drawn from a high-availability beta distribution, and reliability is computed from log-normal MTBF samples through~(5). Structural signatures are generated from heterogeneous feature vectors that encode differences in placement domain, implementation lineage, and software stack, after which the pairwise dissimilarity matrix $D$ is derived. Accordingly, the dataset is synthetic but not arbitrary; it is parameterized to reflect the variability and dependency structure expected in representative cloud-native O-RAN and NFV environments.

The parameter values reported in Tables~\ref{tab:fss_params},~\ref{tab:arq_params}, and~\ref{tab:mldi_params} are selected to provide a balanced and interpretable proof-of-concept setting rather than to optimize a specific deployment. Thresholds, mission horizons, and weighting factors are intentionally kept within moderate ranges so that the reported metrics expose nontrivial changes in structural diversity, admissible substitution, and cross-layer buffering. The compact seven-node visualizations are retained because they keep the substitution and failure trajectories readable in a two-column format. Likewise, the five-algorithm matrix in Fig.~\ref{fig:arq_heatmaps} is used only as an illustrative pairwise example, whereas the main ARQ attack study already uses the expanded 12-algorithm set listed in Table~\ref{tab:arq_params}. Unless otherwise stated, each reported point is averaged over repeated realizations drawn from the same generative model.

\subsubsection{FSS: Functional Substitution Behavior}

\begin{table}[h]
\centering
\caption{Main FSS parameters used in the reported experiment.}
\label{tab:fss_params}
\begin{tabular}{ll}\toprule
Parameter & Value / Description \\\midrule
$\delta$, $A_{\min}$, $R_{\min}$ & $0.5$, $0.5$, $0.5$ \\
$T$, Weighting mode & $168$\,h, Joint (Availability $\times$ Reliability) \\
$q$ list, Trials per $q$ & $\{0,0.05,0.10,0.20,0.30,0.40,0.50\}$, $10$ \\
Attack type & Targeted \\
Functions, Elements & $F1,F2,F3$; $7$ nodes ($e_1$--$e_7$) \\\bottomrule
\end{tabular}
\end{table}

\begin{figure}[h]
    \centering
    \subfloat[Ablation between functional redundancy and baseline FSS. Higher redundancy does not necessarily imply higher structural diversity.]{\includegraphics[width=0.48\columnwidth]{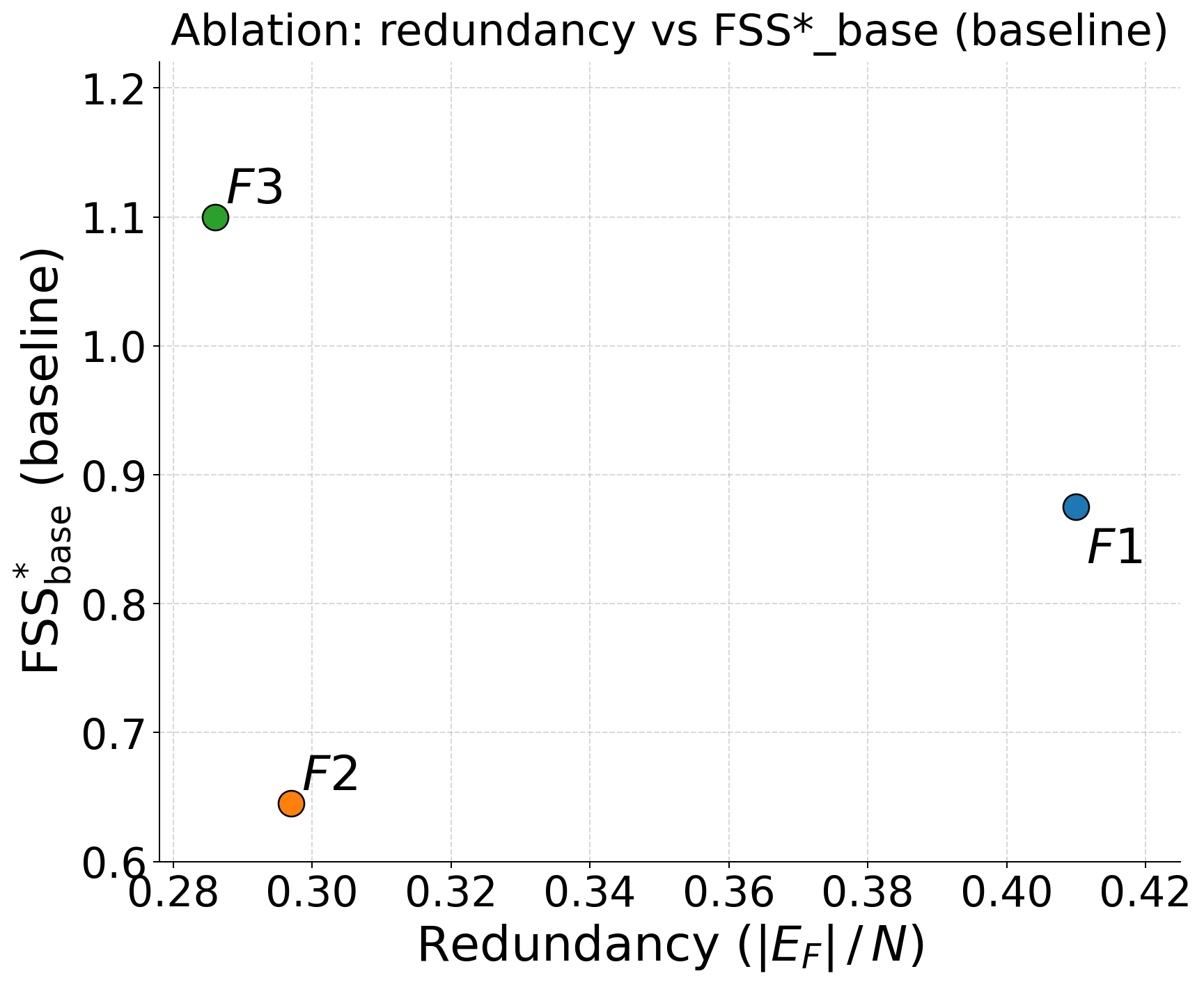}\label{fig:fss_ablation}}
    \hfill
    \subfloat[FSS failure behavior for $F1$ under targeted removals. The baseline score $\mathrm{FSS}^*_{\mathrm{base}}$ collapses sharply beyond moderate disruption, while the weighted score $\mathrm{FSS}^{*(w)}$ remains suppressed due to strict admissibility constraints.]{\includegraphics[width=0.48\columnwidth]{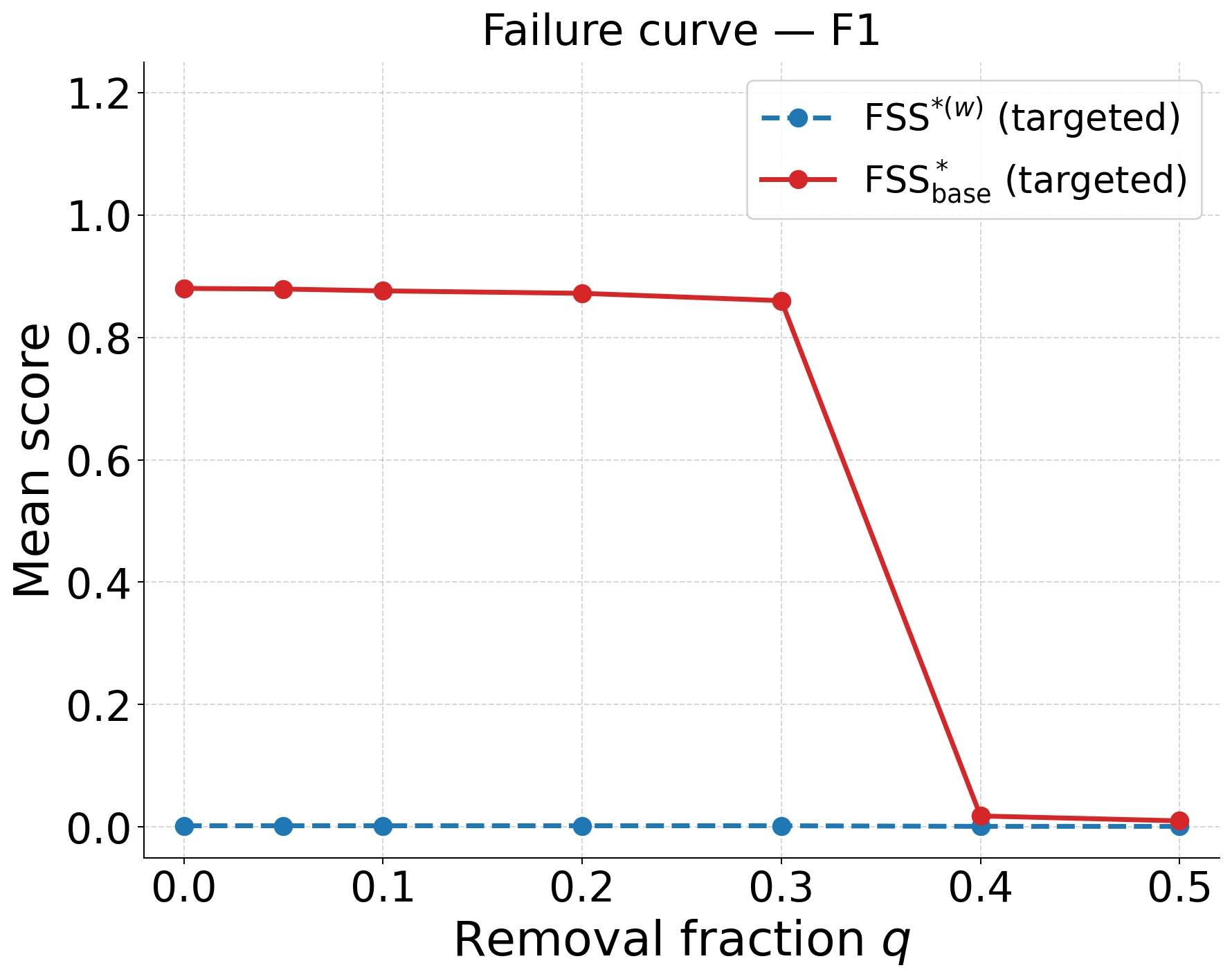}\label{fig:fss_failure}}
    \\
    \subfloat[SoTA comparison under targeted correlated failures. The proposed FSS\,+\,ARQ\,+\,MLDI framework sustains significantly higher functional service continuity compared to standard redundancy and Multi-RAT baselines, with the degeneracy gain most pronounced at moderate-to-high removal fractions.]{\includegraphics[width=0.8\columnwidth]{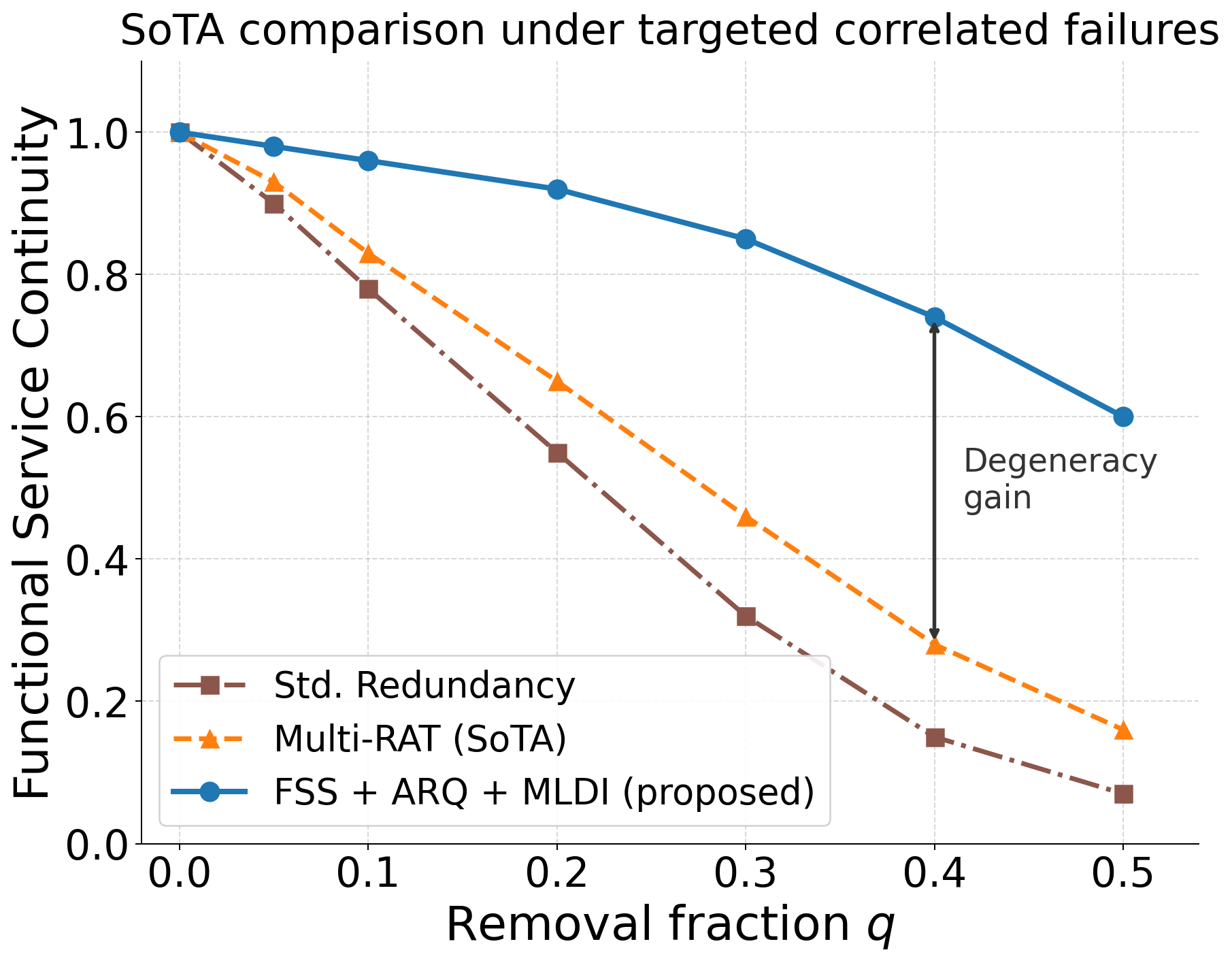}\label{fig:sota}}
    \caption{FSS analysis and SoTA comparison under targeted correlated failures.}
    \label{fig:fss_sota_results}
    \vspace{-3mm}
\end{figure}

We first evaluate component-level resilience using FSS. Fig.~\ref{fig:fss_ablation} shows that redundancy and FSS are not monotonic: although $F1$ has the highest redundancy, its FSS remains below that of $F3$, which achieves the strongest structural diversity with fewer realizations. $F2$ is low in both redundancy and FSS, indicating limited substitution capability. This confirms that replica count alone is not a reliable proxy for resilience; the structural spread of realizations is the more decisive quantity. Fig.~\ref{fig:fss_failure} shows the removal dynamics for $F1$. The baseline score $\mathrm{FSS}^{\star}_{\mathrm{base}}$ remains stable for $q\leq 0.3$, then collapses sharply between $q=0.3$ and $q=0.4$, indicating erosion of structural diversity rather than node count alone. In contrast, $\mathrm{FSS}^{\star(w)}$ stays close to zero throughout because joint availability--reliability filtering excludes structurally distinct but operationally weak substitutes. The separation between the two curves highlights the difference between nominal and practically admissible substitution.

A brief sensitivity check clarifies the role of the hard thresholds in FSS. The baseline score \( \mathrm{FSS}(f) \) is monotone non-increasing in \( \delta \), since increasing the structural threshold can only remove pairs counted as distinct. The weighted score \( \mathrm{FSS}^{\star (w)}(f) \) is additionally monotone non-increasing in \( A_{\min} \) and \( R_{\min} \) under fixed \( A_i \) and \( R_i(T) \), because tighter admissibility conditions can only filter out candidates and cannot create new substitutes. Hence, increasing \( \delta \) from \(0.5\) to \(0.6\) lowers the FSS of \(F_1\), but it does not imply automatic collapse; a sharp collapse occurs only if a substantial fraction of the pairwise dissimilarities for \(F_1\) lies close to the decision boundary. Likewise, increasing \( A_{\min} \) or \( R_{\min} \) suppresses the weighted score more strongly than the baseline score because structurally distinct but operationally marginal substitutes are excluded first. This behavior indicates that the framework is selective rather than fragile: functions supported by broadly separated substitutes remain comparatively stable, whereas functions whose alternatives cluster near the admissibility boundary are more sensitive to threshold tightening.

Fig.~\ref{fig:sota} places the proposed framework against standard redundancy and Multi-RAT baselines under the same targeted attack. Standard redundancy collapses fastest as correlated failures simultaneously eliminate structurally identical replicas. Multi-RAT diversity extends survivability by spreading load across radio interfaces, but without structural decorrelation it still degrades substantially as $q$ increases. The combined FSS\,+\,ARQ\,+\,MLDI design sustains the highest functional service continuity across all removal fractions, with the degeneracy gain widening most sharply beyond $q=0.3$, precisely where redundancy-based schemes have already collapsed.

\subsubsection{ARQ: Pairwise Similarity Structure and Robustness Interpretation}

\begin{table}[h]
\centering
\caption{Main ARQ parameters used in the reported experiment.}
\label{tab:arq_params}
\begin{tabular}{ll}\hline
\textbf{Parameter} & \textbf{Value / Description} \\\hline
$\epsilon$, $\delta$, $\sigma$ & $0.6$, $0.2$, $0.5$ \\
$q$ list, Trials per $q$ & $\{0,0.05,0.10,0.20,0.30,0.40,0.50\}$, $50$ \\
Attack type, Metric & Targeted, ARQ$^{\star}$ heatmap \\
Algorithms, Functions & $A_1$--$A_{12}$; $F1,F2,F3$ \\\hline
\end{tabular}
\end{table}

\begin{figure}[h]
\centering
\includegraphics[width=0.75\linewidth]{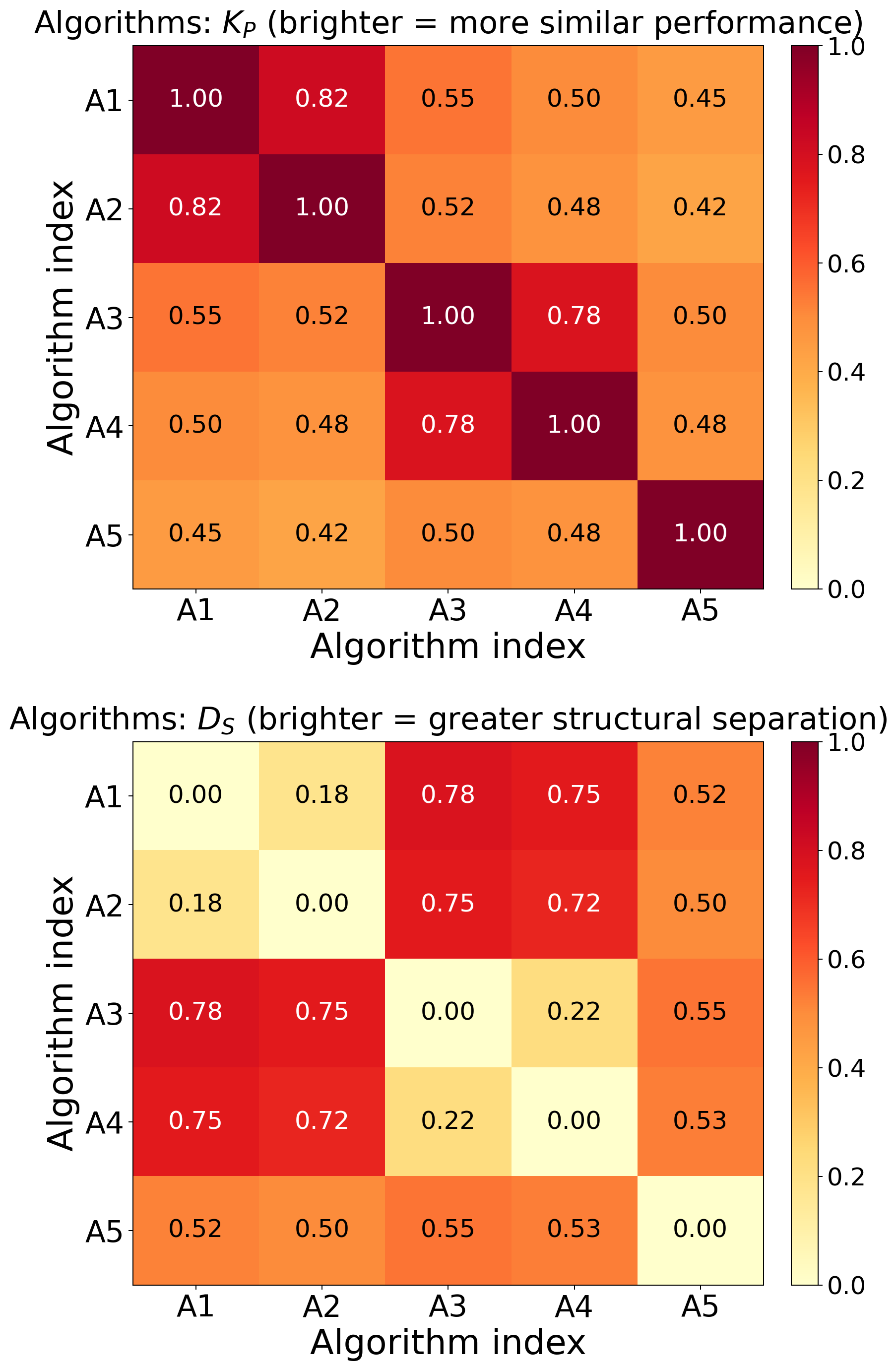}
\caption{ARQ simulator output for the five-algorithm example under $\epsilon=1$, $\delta=0.5$, $\sigma=1.0$. Top: functional similarity $K_P$ (brighter = more similar performance). Bottom: structural dissimilarity $D_s$ (brighter = greater structural separation).}
\label{fig:arq_heatmaps}
\end{figure}

\begin{figure}[h]
\centering
\includegraphics[width=0.9\linewidth]{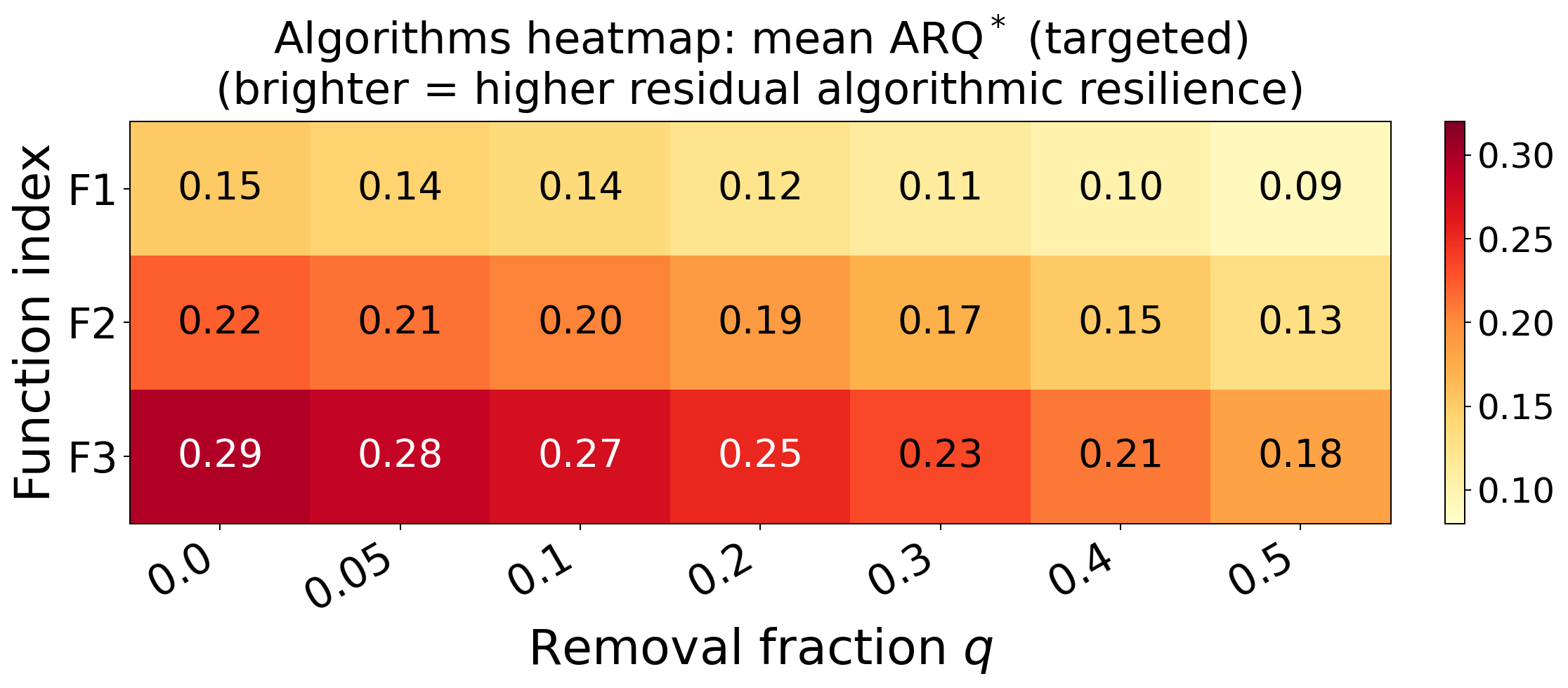}
\caption{Function-level attack heatmap for $\mathrm{ARQ}^{\star}$ under targeted removals. Brighter cells indicate higher mean $\mathrm{ARQ}^{\star}$. $F3$ retains the strongest substitution capacity, $F2$ remains moderate, and $F1$ is comparatively weaker under progressive removals.}
\label{fig:arq_attack}
\end{figure}

We next evaluate algorithmic resilience using ARQ. The simulator reports $\mathrm{ARQ}=0.00000$ and $\mathrm{ARQ}^{\star}=0.279876$. The zero hard score confirms that no pair simultaneously satisfies the strict functional-similarity and structural-separation thresholds, whereas the positive soft score indicates that partially admissible substitutes persist when similarity is treated continuously.

Fig.~\ref{fig:arq_heatmaps} clarifies this behavior. The pairs $(A_1,A_2)$ and $(A_3,A_4)$ are highly similar in performance but exhibit limited structural separation, so they contribute weakly to $\mathrm{ARQ}^{\star}$ and provide weaker resilience under correlated faults than performance similarity alone would suggest. The dominant contribution to $\mathrm{ARQ}^{\star}$ comes from cross-family pairs that preserve acceptable performance similarity while remaining structurally distinct. Algorithm $A_5$ plays an intermediate role by widening the feasible substitution space without dominating the strongest pairwise contributions. Fig.~\ref{fig:arq_attack} extends this view to targeted removals. $F3$ retains the highest residual algorithmic resilience across all removal levels, $F2$ stays moderate, and $F1$ degrades earliest. Since targeted removals prioritize algorithms with high kernel centrality, the most influential performance-compatible bridges are removed first, accelerating collapse for structurally homogeneous portfolios. The design implication is clear: algorithmic resilience is improved not by increasing algorithm count, but by maintaining portfolios of structurally diverse near-equivalent alternatives.

\subsubsection{MLDI: Results and Cross-Layer Interpretation}

\begin{table}[h]
\centering
\caption{Main MLDI parameters used in the reported experiment.}
\label{tab:mldi_params}
\begin{tabular}{ll}\toprule
Parameter & Value / Description \\\midrule
$m$, $k$, $\gamma$ & $4$, $3$, $0.5$ \\
$q$ list, Trials per $q$ & $\{0,0.05,0.10,0.20,0.30,0.40,0.50\}$, $10$ \\
Attack type & Targeted \\
Layers & L1 (physical), L2 (control), L3 (service) \\
Metric outputs & MLDI, MLDI$^{\star}$ \\\bottomrule
\end{tabular}
\end{table}

\begin{figure}[h]
\centering
\includegraphics[width=0.8\linewidth]{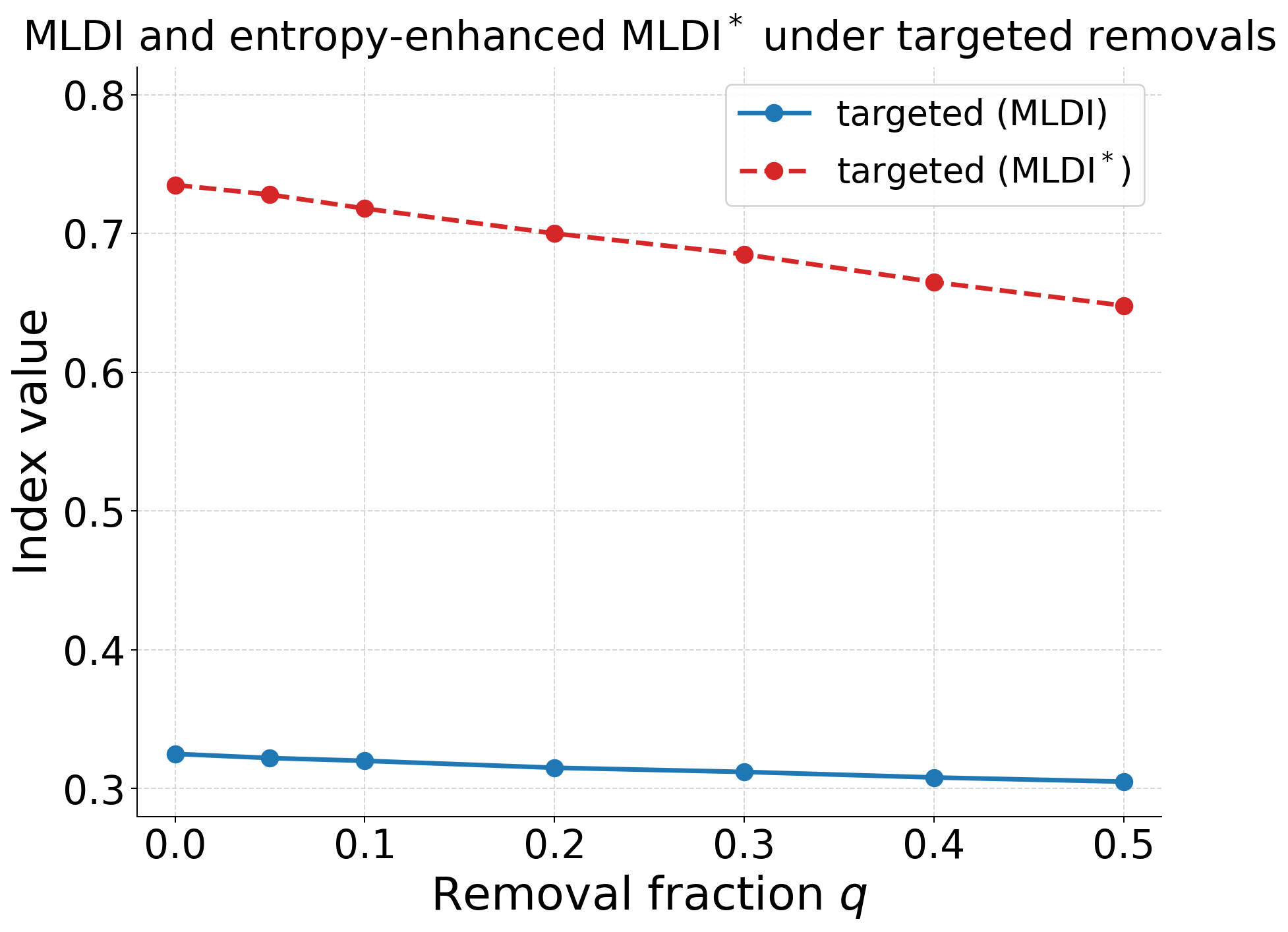}
\caption{MLDI and entropy-enhanced $\mathrm{MLDI}^{\star}$ under targeted removals. The entropy-aware formulation remains consistently higher, indicating sustained functional diversity across layers.}
\label{fig:mldi_line}
\end{figure}

\begin{figure}[h]
\centering
\includegraphics[width=0.9\linewidth]{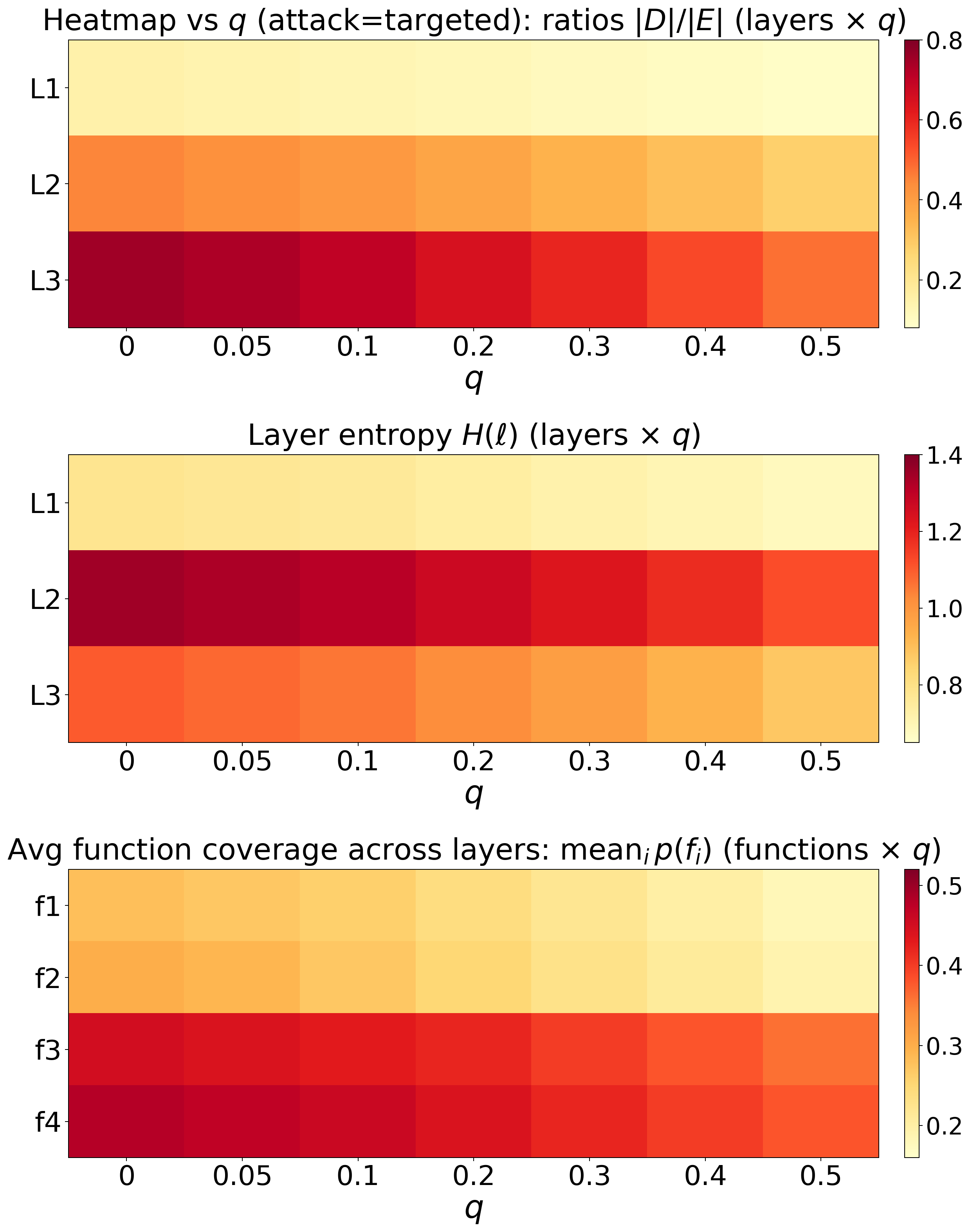}
\caption{Layer-wise behavior under targeted removals. Top: degeneracy ratios $|D_\ell|/|E_\ell|$. Middle: entropy $H(\ell)$. Bottom: average function coverage across layers.}
\label{fig:mldi_heatmaps}
\end{figure}

We next evaluate cross-layer resilience using MLDI. Fig.~\ref{fig:mldi_line} shows that the baseline $\mathrm{MLDI}$ remains relatively low and changes only slightly across all removal fractions, whereas the entropy-enhanced $\mathrm{MLDI}^{\star}$ stays consistently higher and degrades more gradually. This indicates that resilience depends not only on the number of admissible elements, but also on how functionality is distributed across them — a distinction that the entropy formulation captures and the ratio-based baseline misses. Fig.~\ref{fig:mldi_heatmaps} refines this interpretation across three panels. In the top panel, L1 remains dark across all removal levels, indicating strong rigidity and limited substitution capability, while L3 remains brightest, reflecting richer admissible diversity; L2 stays intermediate. In the entropy panel, L2 is consistently brightest, indicating the most balanced functional distribution, whereas L1 remains lowest, confirming that lower layers are both structurally constrained and functionally concentrated. This explains why $\mathrm{MLDI}^{\star}$ remains well above the baseline MLDI. In the bottom panel, $f_3$ and $f_4$ maintain stronger coverage across most removals, whereas $f_1$ and $f_2$ vary more, confirming that resilience is also function-dependent.

\section{Conclusion}

This paper presented a degeneracy-aware framework for resilience analysis in virtualized 6G systems under correlated failures. Through FSS, ARQ, and MLDI, the framework captures substitute diversity across component, algorithmic, and cross-layer levels. The results show that redundancy and robustness are not equivalent: replica-rich systems may remain structurally brittle, algorithmically monocultural, or cross-layer fragile. Degeneracy-aware metrics reveal these hidden weaknesses and show why structurally diverse alternatives support graceful degradation under targeted disruption. These findings position diversification as a practical resilience-by-design principle for open, disaggregated, and softwarized network architectures.

\section*{Acknowledgment}
This work was supported in part by Taighde \'Eireann -- Research Ireland under Grant 13/RC/2077 P2 and by the US-Ireland R\&D Partnership Programme Project ``Resilient Networks'' under Grant RI-SFI-23/US/3924.

\balance


\begin{thebibliography}{99}

\bibitem{Khaloopour2024}
L. Khaloopour, Y. Su, F. Raskob, T. Meuser, R. Bless, L. Janzen, \emph{et al.}, ``Resilience-by-Design in 6G Networks: Literature Review and Novel Enabling Concepts,'' \emph{IEEE Access}, vol. 12, pp. 155666--155695, 2024, doi: 10.1109/ACCESS.2024.3480275.

\bibitem{Ahmad2023}
I. Ahmad, M. A. Habibi, S. Mumtaz, J. Rodriguez, and Z. Han, ``On the Dependability of 6G Networks,'' \emph{Electronics}, vol. 12, no. 6, p. 1472, 2023, doi: 10.3390/electronics12061472.

\bibitem{Deng2024}
S. Deng, X. Li, S. Dustdar, and A. Iosup, ``Cloud-Native Computing: A Survey From the Perspective of Services,'' \emph{ACM Comput. Surv.}, 2024.

\bibitem{Azadiabad2024}
S. Azadiabad and F. Khendek, ``Dependability of Network Services in the Context of NFV: A Taxonomy and State of the Art Classification,'' \emph{J. Network Syst. Manage.}, 2024, doi: 10.1007/s10922-024-09810-2.

\bibitem{Aldas2023}
S. Aldas and A. Babakian, ``Cloud-Native Service Mesh Readiness for 5G and Beyond,'' \emph{IEEE Access}, vol. 11, pp. 132286--132295, 2023, doi: 10.1109/ACCESS.2023.3335994.

\bibitem{Mimran2022}
D. Mimran and Y. Shavitt, ``Security of Open Radio Access Networks,'' \emph{Computers \& Security}, vol. 121, p. 102830, 2022, doi: 10.1016/j.cose.2022.102830.

\bibitem{NIS2022}
NIS Cooperation Group, ``Report on the Cybersecurity of Open Radio Access Networks (Open RAN),'' European Union, May 2022.

\bibitem{NTIA2023}
U.S. National Telecommunications and Information Administration, ``Open RAN Security Report,'' 2023.

\bibitem{Edelman2001}
G. M. Edelman and J. A. Gally, ``Degeneracy and Complexity in Biological Systems,'' \emph{Proc. Natl. Acad. Sci. USA}, vol. 98, no. 24, pp. 13763--13768, 2001, doi: 10.1073/pnas.231499798.

\bibitem{Azariah2024}
W. Azariah, M. A. Ahmad, H. Gacanin, and A. Imran, ``A Survey on Open Radio Access Networks: Challenges, Opportunities, and Future Research Directions,'' \emph{Sensors}, vol. 24, no. 3, p. 1038, 2024, doi: 10.3390/s24031038.

\bibitem{Aldossari2025}
A. Aldossari, M. E. A. Hamza, and M. S. Hossain, ``Reliability and Availability in Virtualized Networks: A Survey on Standards, Modeling Approaches and Research Challenges,'' \emph{arXiv preprint arXiv:2503.22034}, 2025.

\bibitem{DeAlwis2024}
C. De Alwis, P. Porambage, K. Dev, T. R. Gadekallu, and M. Liyanage, ``A Survey on Network Slicing Security: Attacks, Challenges, Solutions and Research Directions,'' \emph{IEEE Commun. Surveys Tuts.}, vol. 26, no. 1, pp. 534--570, 2024, doi: 10.1109/COMST.2023.3312349.

\bibitem{Saeed2025}
M. M. Saeed, \emph{et al.}, ``A Comprehensive Survey on 6G-Security: Physical, Connection and Service Layers,'' \emph{Discover Internet of Things}, 2025.

\bibitem{Failure2024}
``Failure Analysis in Next-Generation Critical Cellular Communication Infrastructures,'' \emph{arXiv preprint arXiv:2402.04448}, 2024.

\bibitem{ETSI2012}
ETSI, ``Network Functions Virtualisation (NFV): Architectural Framework,'' ETSI GS NFV 002, 2012.

\bibitem{Paul2025}
S. Paul, T. Keluskar, and M. Vutukuru, ``A Scalable and Fault-Tolerant 5G Core on Kubernetes,'' 2025.

\bibitem{Barrachina2022}
S. Barrachina-Mu\~noz, \emph{et al.}, ``Cloud-Native 5G Experimental Platform With Over-the-Air Transmissions: Deployments and Monitoring for 5G and Beyond,'' \emph{arXiv preprint arXiv:2207.11936}, 2022.

\bibitem{Gao2024}
S. Gao, S. Quan, and J. Wu, ``Cloud-Native Plinth: A Platform to Support Containerized 5G Core Networks,'' in \emph{Proc. IEEE 21st Int. Conf. on Mobile Ad-Hoc and Smart Systems (MASS)}, 2024, doi: 10.1109/MASS62177.2024.00080.

\end{thebibliography}
\end{document}